 %%%%%%%%%%%%%%%%%% SU6.TEX %%%%%%%%%%%

\input harvmac
\input epsf.tex
\def\caption#1{{\it
	\centerline{\vbox{\baselineskip=12pt
	\vskip.15in\hsize=4.2in\noindent{#1}\vskip.1in }}}}
\def\pyidk{PHY-9057135}

\def\bar#1{\overline{#1}}

\def\bra#1{\left\langle #1\right|}
\def\ket#1{\left| #1\right\rangle}

\def\half{{\textstyle{1\over2}}} %puts a small half in a displayed eqn
\def\frac#1#2{{\textstyle{#1\over #2}}}

%
%       relations
%
\def\ltap{\ \raise.3ex\hbox{$<$\kern-.75em\lower1ex\hbox{$\sim$}}\ }
\def\gtap{\ \raise.3ex\hbox{$>$\kern-.75em\lower1ex\hbox{$\sim$}}\ }
\def\gl{\ \raise.5ex\hbox{$>$}\kern-.8em\lower.5ex\hbox{$<$}\ }
\def\roughly#1{\raise.3ex\hbox{$#1$\kern-.75em\lower1ex\hbox{$\sim$}}}

\def\eg{\hbox{\it e.g.}}

\def\np#1#2#3{{Nucl. Phys. } B{#1} (#2) #3}
\def\pl#1#2#3{{Phys. Lett. } {#1}B (#2) #3}
\def\prl#1#2#3{{Phys. Rev. Lett. } {#1} (#2) #3}
\def\physrev#1#2#3{{Phys. Rev. } {#1} (#2) #3}

\relax
\def\Dsl{\,\raise.15ex \hbox{/}\mkern-13.5mu D}
\def\[{\left[}
\def\]{\right]}
\def\({\left(}
\def\){\right)}
\noblackbox
%\draftmode
\def\pyidk{PHY-9057135}
\Title{\vbox{
\hfill DOE/ER/40561-230-INT95-00-104  \smallskip
\hfill UW/PT 95-14  \smallskip
\hfill CMU-HEP95-13  \smallskip
\hfill DOE-ER/40682-102}}
{\vbox{\vskip-.5in\centerline{The Spin-Flavor Dependence of Nuclear
Forces}\vskip.2in\centerline{
{}From Large-$N$ QCD}}}
\vskip-.1in
\centerline{David B. Kaplan}
\centerline{{\sl
Institute for
Nuclear Theory, University of Washington}}
\centerline{{\sl Box 351550,
Seattle WA 98195-1550, USA}}
\centerline{{\tt dbkaplan@phys.washington.edu}}
\medskip
\centerline{{\it and}}
\medskip
\centerline{Martin J. Savage
\footnote{$^{\dagger}$}{DOE Outstanding Junior Investigator}}
\centerline{{\sl Department of Physics, Carnegie Mellon University}}
\centerline{{\sl
Pittsburgh PA 15213, USA}}
\centerline{
{\tt savage@thepub.phys.cmu.edu}}
\bigskip\bigskip\vfill
{  We show that nuclear interactions are $SU(4)$  symmetric
at leading order in chiral perturbation theory in
the large-$N$ limit of QCD.
The nucleons and  delta resonances form a
20-dimensional representation of $SU(4)$ and
we show how Wigner's supermultiplet symmetry  $SU(4)_{sm}$, under which the
nucleons transform as a 4-dimensional representation,
follows as an accidental low energy symmetry.
Exploiting $SU(4)$ symmetry allows one to express the 18
independent leading $N$, $\Delta$ interaction operators
invariant under $SU(2)_I\times SU(2)_J$ in terms of only two couplings.
The three flavor analogue allows one to express the 28
leading octet, decuplet interactions in terms of only two
couplings, which has implications for  hypernuclei and
strangeness in ``neutron'' stars. }
\Date{9/95}
\baselineskip 18pt

\newsec{Implications of spin-flavor symmetry in effective nuclear forces}

Short distance nuclear forces relevant for low energy processes can be
incorporated into chiral Lagrangians in terms of local operators in a
derivative expansion \nref\weinberg{
 S. Weinberg, \pl{251}{1990}{288};
\np{363}{1991}{3}; \pl{295}{1992}{114}}\nref\kolck{
C. Ordonez, U. van Kolck, \pl{291}{1992}{459};
C. Ordonez, L. Ray, U. van Kolck, \prl{72}{1994}{1982}; U. van Kolck,
\physrev{C49}{1994}{2932}}\refs{\weinberg,\kolck}.  There are two leading
(dimension six) operators involving nucleons alone, given by
\eqn\csct{\CL_{6}= - \half C_S (N^{\dagger}N)^2 - \half C_T (N^{\dagger}
\vec\sigma N)^2}
where $N$ are isodoublet two-component spinors, and the $\vec\sigma$ are Pauli
matrices.  Higher derivative operators account for the spin-orbit coupling,
among other effects\foot{In low energy nucleon-nucleon scattering the higher
derivative terms will be less important than the leading operator.  However,
many-body effects in large nuclei can enhance the importance of subleading
operators, such as the spin-orbit interaction.}.  Including the $\Delta$
isobars in the theory leads to 18 independent dimension six operators allowed
by spin and isospin symmetry\foot{It is simplest to count operators in the form
$(\psi_1\psi_2)(\psi_3\psi_4)^{\dagger}$, requiring $(\psi_1\psi_2)$ and
$(\psi_3\psi_4)$ to have the same spin and isospin quantum numbers.  One finds
the above two $(NN)(NN)^{\dagger}$ operators;  zero operators of the form
$(NN)(N\Delta)^{\dagger}$; two $(NN)(\Delta\Delta)^{\dagger}$, four
$(N\Delta)(N\Delta)^{\dagger}$, two
$(N\Delta)(\Delta\Delta)^{\dagger}$, and eight
$(\Delta\Delta)(\Delta\Delta)^{\dagger}$ operators. }.  In order to discuss
hypernuclei, or strangeness in dense matter,   one must consider $SU(3)$ flavor
multiplets --- there are six independent leading operators involving the baryon
octet alone \ref\sav{
M. J. Savage, M. B. Wise, hep-ph/9507288
}, while including the decuplet inflates the number to 28 independent
operators. The number of independent dimension seven interactions is still
much greater.

Clearly, to make headway in a systematic effective field theory analysis of
nuclear and hypernuclear forces, it is desirable to find some simplifying
principle.  In this letter we propose that among the baryon interactions,
$SU(4)$ spin-flavor symmetry for two flavors, or $SU(6)$ symmetry for three
flavors should  be a good approximation.  We show how these symmetries have a
vastly simplifying effect on the dimension six interactions described above,
reducing both the 18 $N-\Delta$ interactions and the 28 octet-decuplet
interactions down to just two independent operators.  We support our allegation
that spin-flavor symmetry is relevant to nuclear forces first by outlining
its implications and by giving   empirical evidence in support of $SU(4)$ in
nuclei.  Then   we prove that these symmetries become exact in the large-$N$
limit of QCD.

Under $SU(2f)$ symmetry  the two spin states and $f$ flavors of quarks
transforming as the $2f$ dimensional defining representation.  For $N=3$, the
lowest lying
baryons have the quantum numbers  of three quarks in an $S$-wave, transforming
as a three index
symmetric tensor $\Psi^{\mu\nu\rho}$ under $SU(2f)$.  For $f=2$, $\Psi$ is the
20-dimensional representation of $SU(4)$, comprising of the four $N$ and
sixteen $\Delta$ spin/isospin states; for $f=3$, $\Psi$ is the 56-dimensional
$SU(6)$ representation    containing the $J=\half$ octet and the
$J=\frac{3}{2}$ decuplet.  In either case one finds that there are only two
$SU(2f)$ invariant dimension six operators.  These can be written in terms of
the baryon fields $\Psi$ as
\eqn\suivop{\CL_{6} = -{1\over f_\pi^2}\[ a (\Psi^{\dagger}_{\mu\nu\rho}
\Psi^{\mu\nu\rho})^2 +  b
\Psi^{\dagger}_{\mu\nu\sigma} \Psi^{\mu\nu\tau}\Psi^{\dagger}_{\rho\delta\tau}
\Psi^{\rho\delta\sigma}\]\ ,}
where $f_\pi=132$ MeV is the pion decay constant.

Eq. \suivop\  can be expressed in terms of the more familiar fields by writing
 each $SU(2f)$ index $\mu$  as a pair of  flavor and spin indices $(i\alpha)$
under $SU(f)\times SU(2)_J$, and then by projecting out components with the
desired $SU(f)\times SU(2)_J$ transformation properties.  In this way one finds
for two flavors
\eqn\decompii{
\Psi^{(\alpha i)(\beta j)(\gamma k)} = \Delta^{ijk}_{\alpha\beta\gamma} +
{1\over\sqrt{18}} \(N^i_\alpha \epsilon^{jk}\epsilon_{\beta\gamma} + N^j_\beta
\epsilon^{ik}\epsilon_{\alpha\gamma}+
N^k_\gamma\epsilon^{ij}\epsilon_{\alpha\beta}\)\ ,}
and for three flavors
\eqn\decompiii{\Psi^{(\alpha i)(\beta j)(\gamma k)}=
T^{ijk}_{\alpha\beta\gamma} + {1\over\sqrt{18}}\(
B^i_{m,\alpha}\epsilon^{mjk}\epsilon_{\beta\gamma} +
B^j_{m,\beta}\epsilon^{mki}\epsilon_{\gamma\alpha} +
B^k_{m,\gamma}\epsilon^{mij}\epsilon_{\alpha\beta} \) .}
In the above expressions $N$, $\Delta$, $B$, and $T$ are the nucleon, isobar,
octet and decuplet fields respectively.  Indices $i,j\ldots$ denote flavor
indices while $\alpha,\beta\ldots$ are spin indices. $\Delta$ and $T$ are
totally symmetric tensors separately in flavor and spin; $B$ is a traceless
matrix in flavor.  The normalization factors of $1/\sqrt{18}$ are fixed so that
the baryon number operator equals $(\Psi^{\dagger}_{\mu\nu\rho}
\Psi^{\mu\nu\rho})$.

 By plugging expressions \decompii, \decompiii\ into our $SU(2f)$ symmetric
Lagrangian \suivop\ it is possible to determine all of the leading short range
interactions between the $J=\half$ and $J=\frac{3}{2}$ baryons in terms of the
two coefficients $a$ and $b$.  We spare the reader all of the results, and
focus on predictions for interactions solely involving the $J=\frac{1}{2}$
baryons. For two flavors,  the $SU(4)$ symmetry yields predictions for the
Weinberg coefficients of eq. \csct\ in terms of $a$ and $b$:
\eqn\resii{C_S={2 (a-b/27) \over f_\pi^2}\ ,\qquad C_T=0\qquad (two\ flavors).}

For three flavors, $SU(6)$ symmetry predicts the six Savage-Wise (SW)
coefficients $c_1\ldots c_6$ \sav\ for the interactions involving four baryon
octet fields:
\eqn\resiii{
\eqalign{c_1&= -\frac{7}{27} b\cr c_2&= \frac{1}{9}b\cr c_3&= \frac{10}{81}b
\cr}\qquad
\eqalign{c_4&= -\frac{14}{81}b\cr c_5&= a+\frac{2}{9}b \cr c_6&= -
\frac{1}{9}b\cr}\qquad\quad
(three\ flavors)\ .}
 Given that $C_S=(2c_1+c_2+2c_5+c_6)/f_\pi^2$ and
$C_T=(c_2+c_6)/f_\pi^2$, the
$SU(6)$ prediction \resiii\ contains the $SU(4)$ prediction \resii.
Note that the contributions proportional to $b$ are quite suppressed.
In fact, as we discuss below, in the large-$N$ limit of QCD,
both $a/f_\pi^2$ and $b/f_\pi^2$ are $\CO(N)$, while there are
$\CO(1)$ ($\sim 30\% $) violations to the $SU(6)$ predictions.
Similarly one expects $\sim 30\% $ violations due to $SU(3)$ breaking by
the strange quark mass.
Thus the contributions proportional to $b$ in the above
relations are subleading, and it is consistent to take $b=0$ in eq. \resiii\
to leading order in $N$. Note that with the $b$ terms negligible,
the self-interactions  among the  $J=\half$ octet baryons
are invariant under an accidental $SU(16)$ symmetry.  For two flavors, the
prediction $C_T=0$ implies an accidental $SU(4)$ symmetry among the nucleon
self-interactions.

\newsec{Evidence for $SU(4)$ symmetry in nuclear physics}

Why should one believe that there is an approximate $SU(2f)$ symmetry in the
short-range nuclear forces?  We first give empirical evidence for $SU(4)$
symmetry in nuclear physics; we then show that in the large-$N$ limit of QCD,
$SU(4)$ relations are accurate to order $1/N^2$ ($\sim 10\%$), while $SU(6)$
relations are accurate to $\CO(1/N)$ ($\sim 30\%$) and $\CO(m_s)$.

The testable consequence of $SU(4)$ symmetry is the prediction \resii\ that
$C_T=0$, which implies the existence of an {\it accidental} symmetry in low
energy nucleon interactions.  This symmetry is Wigner's ``supermultiplet
symmetry'' \ref\wigner{E. Wigner, \physrev{51}{1937}{106}, 947; {\it ibid.} 56
(1939) 519}, which we will denote $SU(4)_{sm}$; it arises because with $C_T=0$
the interaction \csct\ is simply proportional to $(N^{\dagger}N)^2$ which is
invariant under an $SU(4)$ with the four spin and isospin nucleon states
transforming as a 4-dimensional representation.  Wigner's $SU(4)_{sm}$  is not
equivalent to the more fundamental $SU(4)$ symmetry,
under which the nucleon transforms as part of a 20-dimensional representation
along with the deltas.  Since
$SU(4)_{sm}$ is broken by dimension seven operators (\eg\ spin-orbit
interactions) as well as dimension six $N-\Delta$ interactions, it is  only
expected to be an approximate symmetry in light nuclei.

The validity of $SU(4)_{sm}$ in light nuclei is supported by an array of
evidence.  An extensive discussion is given in ref.
\ref\parikh{{\it Group Symmetries in Nuclear Structure}, J.C. Parikh, Plenum
press, (1978)} ; additional support can be found in the literature in the
context of  electron scattering
\ref\dw{T.W. Donnelly and G.E. Walker, Ann. of Phys., 60 (1970) 209},
 giant dipole resonance multiplets \ref\wal{{\it Theoretical Nuclear Physics
and Subnuclear Physics}, by J.D. Walecka, Oxford University Press (1995)},
$\beta$-decay selection rules in $A=19$ nuclei \ref\elliot{J.P. Elliott, in
{\it Isospin in Nuclear Physics},
page 73, ed. D.H. Wilkinson, North Holland Publishing, (1969)},
and double $\beta$-decay \ref\dbeta{P. Vogel, M. R. Zirnbauer,
\prl{57}{1986}{3148}}.  A particularly illuminating discussion of $SU(4)_{sm}$
symmetry in nuclei is found in \ref\vogorm{P. Vogel, W.E. Ormand, \physrev
{C47}{1993}{623}}.
We will mention here two particular pieces of evidence for $SU(4)_{sm}$: the
two nucleon system, and $\beta$-decays of $A=18$ nuclei.

The two-nucleon system provides a striking example   of $SU(4)_{sm}$ symmetry
in nuclear interactions.
The $I=0$, $S=1$ channel has a scattering length of $a_1 = 5.423 {\rm fm}$ and
a bound state (the deuteron) with   binding energy of $E_B =  2.225 {\rm MeV}$.
In contrast the $I=1$, $S=0$ channel has  scattering lengths  $a_0\sim -20 {\rm
fm}$, and the threshold states are nearly bound.  In the $SU(4)_{sm}$ limit
both channels would be identical, and so the very different scattering lengths
would make it appear that $SU(4)_{sm}$ is badly broken.
However,  scattering lengths are very sensitive to small changes in the
interaction potential if there are almost bound or almost unbound eigenstates,
as is the case in the nucleon-nucleon system.  Thus there may be an approximate
symmetry in the interaction potentials which is not evident in the scattering
lengths.
Modelling the nuclear two-body potential by a spherically symmetric, finite
depth square well  one finds that the relevant potential parameters ($V_0$ the
depth and $R$ the range)
are $V_{0t} = 38.5 {\rm MeV}$, $R_t = 1.93 {\rm fm}$ for the spin-triplet state
and
 $V_{0s} = 20.3 {\rm MeV}$, $R_s = 2.50 {\rm fm}$ for the spin singlet states
\ref\enp{{\it Elements of Nuclear Physics}, W.E. Burcham, Longman Group Ltd,
(1979)}.
As we will show in the next section,  the large-$N$ analysis predicts that the
interaction should be $SU(4)$ symmetric in the Born approximation; beyond Born
approximation, infrared divergences can mask the result.
If we match the Weinberg coefficients $C_{S,T}$ at scattering momenta
considerably above threshold (but below the scale where the derivative
expansion breaks down) then the Born expansion in the full and effective
theories should match pretty well.  Thus $C_{S,T}$ are proportional to the
spatial integrals of the nucleon-nucleon potential;  in the case of the model
of a spherically symmetric, square well potential described above, this leads
to
\eqn\beudy{\eqalign{
C_S = & -{\pi\over 3} \left(  3 V_{0t} R_t^3 + V_{0s} R_s^3  \right)  \ \ =\ \
-2.24 {1\over f_\pi^2} \cr
C_T = & -{\pi\over 3} \left(  V_{0t} R_t^3 - V_{0s} R_s^3  \right)  \ \ =\ \
0.126 {1\over f_\pi^2}
}\ \ \ .}
Evidently the nuclear two-body interaction is $SU(4)_{sm}$ invariant with
corrections at the
$C_T/C_S\simeq 5\%$ level, consistent with what one might expect from a
$1/N^2$ effect with $N=3$.  Note that this fit fixes one linear combination of
the $a$ and $b$ coefficients, as given in eq. \resii.
Weinberg chose to fit $C_S$ and $C_T$ at threshold, so that they were
proportional to scattering lengths; this procedure yields  $C_T/C_S \sim -4$,
 completely obscuring the approximate $SU(4)$ symmetry \weinberg.

\topinsert
\centerline{\epsfxsize=4.5in\epsfbox{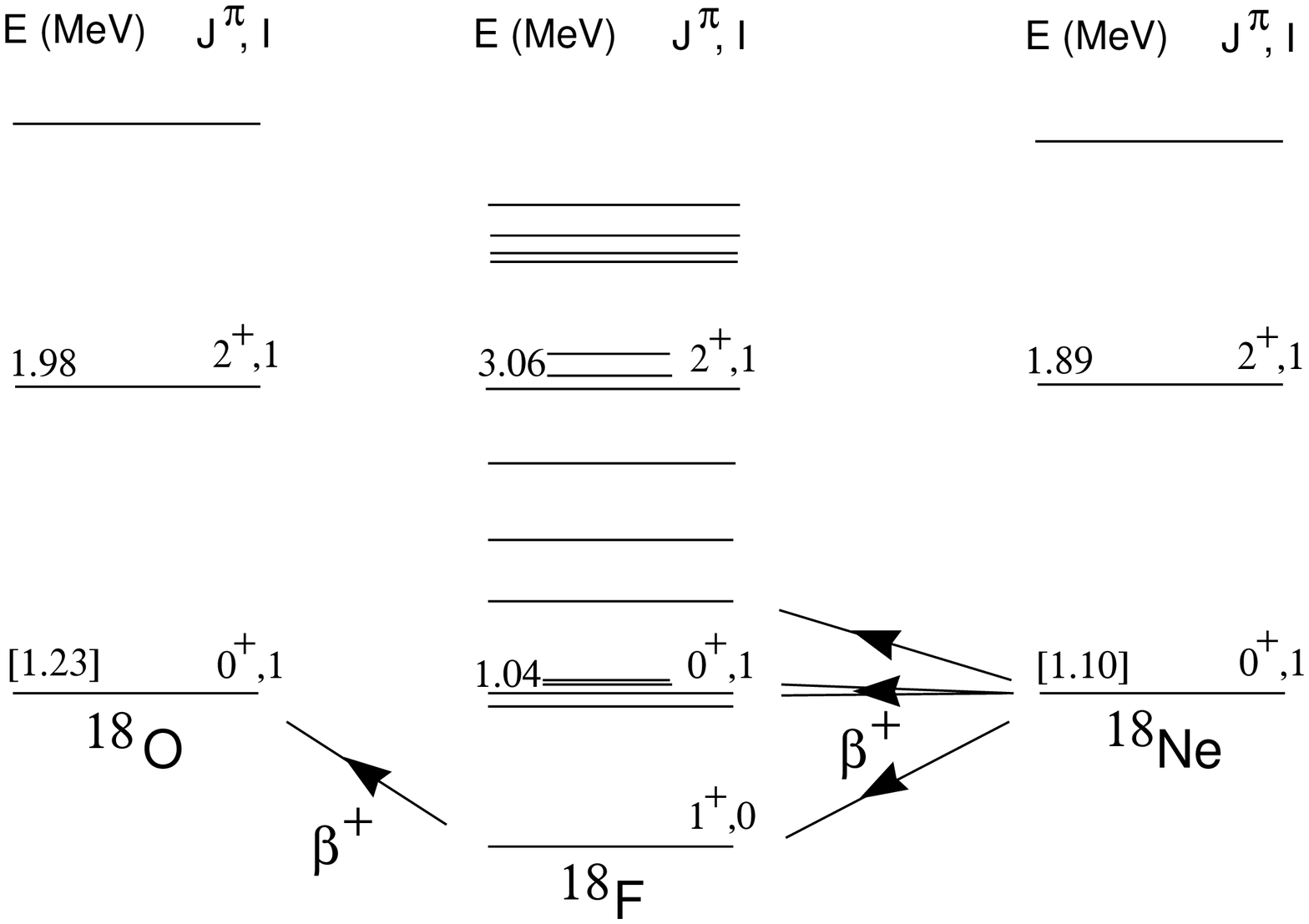}}
\smallskip
\caption{Fig. 1.  Energy level diagram for $^{18}O$ , $^{18}F$ and $^{18}Ne$.
The $n-p$ mass difference and coulomb effects have been removed.
The energies in square brackets  are relative to the $^{18}F$ ground state.
$\beta$-decays to various states are indicated by arrows.}
\endinsert

A second demonstration of $SU(4)_{sm}$ symmetry in nuclear interactions can
be found in the strengths of Gamow-Teller transitions between  light nuclei.
The Gamow-Teller operator is an $SU(4)_{sm}$ generator and as such cannot
induce transitions between states belonging to different irreducible
representations of $SU(4)$.
We have found a particularly compelling demonstration in transitions between
the $A=18$ nuclei, $^{18}O$ , $^{18}F$ and $^{18}Ne$.
The energy level diagram for the $A=18$ nuclei and the
observed $\beta$-decays are shown in Fig. 1,
and can be found in more detail in
\ref\selove{F. Ajzenberg-Selove, Nucl. Phys. A475 (1987) 1}.
The naive structure of these nuclei is two valence nucleons in the $s-d$ shell
on a closed, inert $^{16}O$ core (the predominant mixing with deformed states
is expected to involve four nucleons in an $SU(4)_{sm}$ singlet state in the
$s-d$ shell and two holes in the core, which will not affect our classification
of these states under $SU(4)_{sm}$).
The results of shell-model computations
\nref\bertsch{ {\it The Practitioner's Shell Model}, G.F. Bertsch,
North-Holland/American Elsevier (1972).}\nref\rays{R. Sorensen, {\it private
communication}}\refs{\bertsch,\rays},
show that the two valence nucleons in the
lowest lying $(J^\pi,I) = (0^+,1)$ and $(1^+,0)$ states are predominantly in a
relative $L=0$ configuration. This allows us to identify the ground states of
$^{18}O$ , $^{18}F$ and $^{18}Ne$ and the excited state of $^{18}F$ at $E=1.04
{\rm MeV}$ (as shown in Fig. 1) as the   members of the ${\bf 6}$
representation of $SU(4)_{sm}$. (The $^{18}F$ ground state has $(J,I)=(1,0)$,
while the other three states form a $(J,I)=(0,1)$ multiplet).
In fact, shell-model computations indicate that these states are in the
same $SU(4)_{sm}$ multiplet with a probability of $87\%$
\ref\Georgepriv{G. Bertsch, {\it private communication }}; the ``missing''
13\% is largely due to the spin-orbit interaction.
The $\beta$-decay strength of transitions between states of this supermultiplet
along with the strengths of transitions to states outside the supermultiplet
are shown in Table 1
\ref\adel{E.G. Adelberger {\it et al}, \physrev{C27}{1983}{2833}}.

\topinsert
\vbox{
\input tables
\begintable
Decay |  $(J^\pi,I)\ \rightarrow (J^\pi,I) $ | $\log_{10}(ft) $ | R  \cr
$^{18} F {\buildrel \beta^+ \over \longrightarrow} {^{18}O} (g.s)  $
| $(1^+,0)\rightarrow (0^+,1)$ | $3.554$ | 0.73 \crnorule
$^{18} Ne {\buildrel \beta^+ \over \longrightarrow} {^{18}F} (g.s) $ |
$(0^+,1)\rightarrow (1^+,0)$ | $3.096\pm 0.004$ | 0.71 \crnorule
$^{18} Ne {\buildrel \beta^+ \over \longrightarrow} {^{18}F} (1.04 {\rm MeV})
$
| $(0^+,1)\rightarrow (0^+,1)$ | $3.473\pm 0.013$ | 1\crnorule
$^{18} Ne {\buildrel \beta^+ \over \longrightarrow} {^{18}F} (1.08 {\rm MeV})
$
| $(0^+,1)\rightarrow (0^-,0)$ | $7.012\pm 0.059$  | 0.017 \crnorule
 $^{18} Ne {\buildrel \beta^+ \over \longrightarrow} {^{18}F} (1.70 {\rm MeV})
$
| $(0^+,1)\rightarrow (1^+,0)$ | $4.477\pm 0.015$ | 0.15
\endtable}
\caption{Table 1: $\log_{10}(ft) $ values for $\beta$-decay between A=18
nuclei.  The first three decays are between states in a $\bf 6$ of
$SU(4)_{sm}$; the last two are decays between different $SU(4)_{sm}$
multiplets.  $R$ measures the matrix element relative to that for the third
entry (corrected for phase space, final state multiplicity, $g_A$
factors).  }
\endinsert

It is seen from the first three entries of Table 1 that the Gamow-Teller
transitions ($\propto \, \vec\sigma\tau_+$) to states within the
$SU(4)$ supermultiplet are of similar strength as the superallowed transition
($\propto \, \tau_+ $),
$^{18} Ne {\buildrel \beta^+ \over \longrightarrow} {^{18}F} (1.04 {\rm MeV})
$,
despite the fact that they are not related by isospin.
Further, it is clear from the last two entries that the decays to states
outside the supermultiplet are much weaker than those to states within the
supermultiplet.  To be quantitative, we give the ratio of matrix elements
(corrected for Coulomb interactions, $g_A$, and final state multiplicities)
relative to the superallowed transition as the parameter $R$ in the last
column.  Isospin predicts equality between the first two entries; $SU(4)_{sm}$
predicts equality between the first three entries.  Allowing for a 13\% effect
due to contamination of the wavefunctions largely due to the spin-orbit
interaction, we
see that $SU(4)_{sm}$ is a good symmetry in the dimension six interactions to
within $\sim 20\%$, again roughly consistent with $SU(4)_{sm}$ violation at the
$1/N^2$ level \foot{If the contamination is greater than  13\% --- as is
suggested by the fact that the shell model systematically overestimates
Gamow-Teller strengths \ref\shellgt{G. F. Bertsch, H. Esbensen, Rep. Prog.
Phys. 50 (1987) 607} --- then the agreement with with the $SU(4)_{sm}$
prediction may be better than 20\%.}.

\newsec{Spin-flavor symmetry from large-$N$ QCD}

We now turn to a theoretical justification for spin-flavor symmetry in the
leading nuclear forces, appealing to the $1/N$ expansion of QCD.
The internal structure of baryons becomes greatly simplified as that
the number of colors $N$ becomes large.
In this limit, baryons are comprised of $N$ quarks and a baryon mass scales
like $N$.  As shown by 't Hooft, virtual $q\bar q$ pairs are suppressed in this
limit \ref\thooft{G. 't Hooft, \np{72}{1974}{461}; {\it ibid.} B75 (1974) 461};
subsequently  Witten showed that the quarks in a baryon obey a relativistic
Hartree equation \ref\witten{E. Witten, \np{160}{1979}{57}}. For two flavors
one can also prove certain $SU(4)$ relations among operator matrix elements in
single
large-$N$ baryons states \nref\gsak{
J.L. Gervais, B. Sakita, \prl{52}{1984}{87}} \nref\djm{
R. Dashen, A. V. Manohar, \pl{315}{1993}{425};
E. Jenkins, \pl{315}{1993}{441};
E. Jenkins, A. V. Manohar, \pl{335}{1994}{452};
R. Dashen, E. Jenkins, A. V. Manohar, \physrev{D49}{1994}{4713}; {\it ibid.}
D51 (1995) 2489}\nref\djmii{R. Dashen, E. Jenkins, A. V. Manohar,
\physrev{D51}{1995}{3697}}\nref\hg{
C. D. Carone, H. Georgi, S. Osofsky, \pl{322}{1994}{227};
C. D. Carone, H. Georgi, L. Kaplan, D. Morin,
\physrev{50}{1994}{5793}}\nref\lmr{
M. A. Luty, J. March-Russell, \np{426}{1994}{71}}\refs{\gsak-\lmr}.  The
existence of these relations is  due to the fact that while large-$N$ analogues
of the nucleon have $N$ quarks, they still have $\CO(1)$ spin and isospin,
so that the spin-dependent Hartree potential experienced by quarks is down by
$1/N$
relative to the spin independent potential.  With three flavors one finds
$SU(6)$ relations, provided one considers baryons with $\CO(1)$ strange quarks,
as well as $\CO(1)$ spin.   $SU(4)$ relations among 1-baryon matrix elements
are found to work to $\CO(1/N^2)$  \djm, while $SU(6)$ relations get $1/N$
corrections \refs{\djm,\lmr};  some of $1/N$ corrections can be computed,
however \refs{\djm,\djmii}.  We will find similar symmetries and corrections
in the 2-baryon sector.

\topinsert
\centerline{\epsfxsize=2.5in\epsfbox{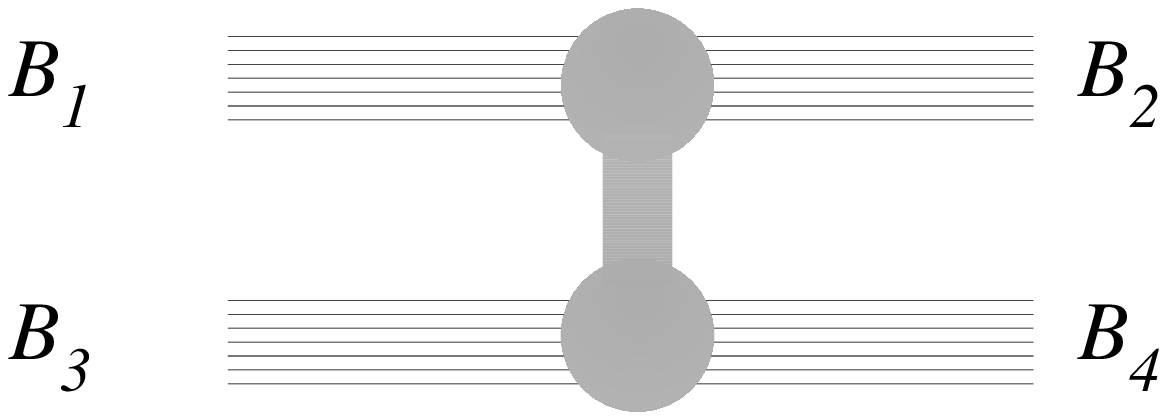}}
\smallskip
\caption{Fig. 2.  A connected baryon-baryon interaction in large-$N$ QCD due to
the exchange of quarks and gluons.}
\endinsert

 Witten discussed how best to consider baryon-baryon collisions \witten: one
works with the time dependent Schrodinger equation in (relativistic) Hartree
approximation.  It is not possible to construct the Hartree potential
explicitly, but we can deduce certain of its properties by making the
reasonable assumption it is given by the sum of connected
Feynman diagrams in a $1/N$ expansion.  Such diagrams have the generic form  of
Fig. 2, where the blob represents the exchange of an arbitrary number of quarks
between the two baryons, as well as gluon exchange between the quarks.

\topinsert
\centerline{\epsfxsize=2.5in\epsfbox{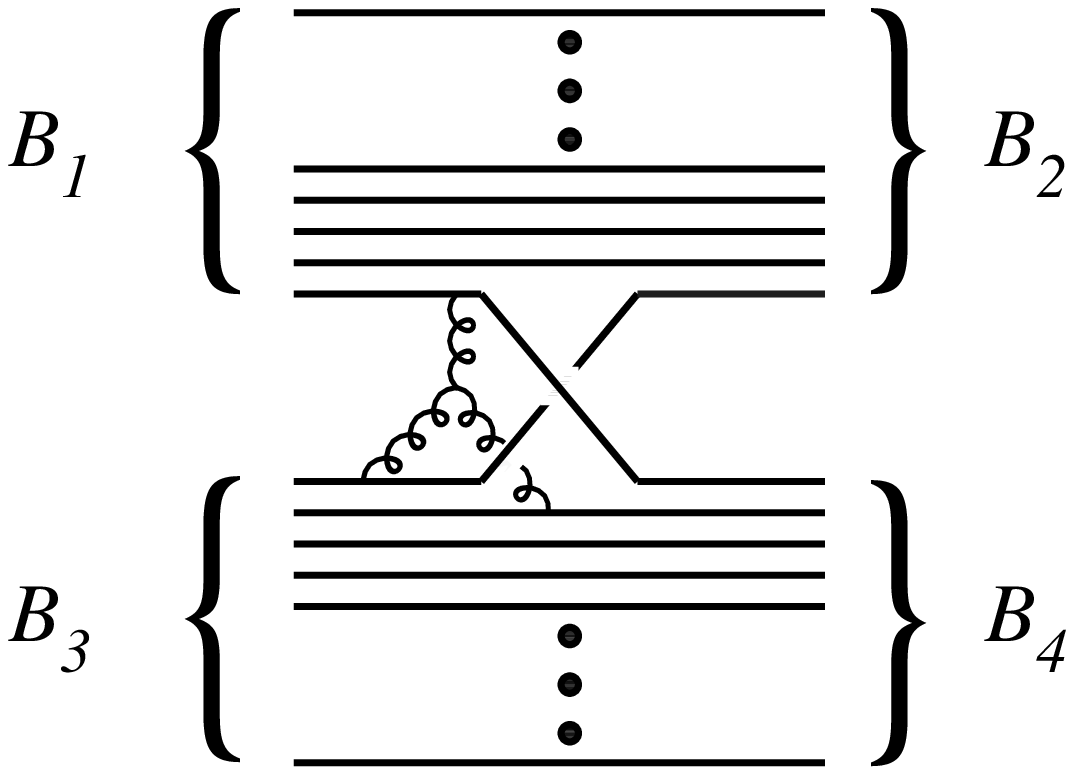}}
\smallskip
\caption{Fig. 3.  A connected diagram that is $\CO(1/N^2)$ involving a 1-body
operator between $\bra{B_1}$ and $\ket{B_2}$, and a 2-body operator between
$\bra{B_3}$ and $\ket{B_4}$.}
\endinsert
 A specific example of such a diagram is given in Fig. 3; it is seen to be
$\CO(1/N^2)$ since each gluon coupling brings a factor of $1/\sqrt{N}$, and
there are no closed color loops in the diagram.  Since there are three quarks
involved and $N$ possible choices for each quark, the matrix element of this
operator will involve a combinatoric factor of $N^3$, making its net
contribution to the baryon-baryon scattering amplitude  $\CO(N)$.  One can
easily generalize to interactions involving any number of quark exchanges:
leading connected graphs involving $r$ quarks scale as $N^{(1-r)}$, and their
contribution to the amplitude is $\CO(N)$ \witten.

Our analysis of the baryon-baryon scattering diagrams will use techniques
and notation similar to  those in \refs{\djm-\lmr} --- most closely those of
ref. \djmii. It
is convenient to classify the connected diagrams contributing to the
interaction in Fig. 2 by the number of quarks $n$ involved on the $B_1-B_2$
baryon line, by the number of quarks $n'$ involved on the $B_3-B_4$ line, and
by
the total isospin and spin $(I,J)$ transmitted between the two baryon lines
\foot{The identity of the two baryon lines can be kept distinct by considering
the number of quarks involved $n$ and $n'$ to be $\CO(1)$; diagrams with $n$ or
$n'$ of $\CO(N)$ are actually suppressed since the combinatoric factor for
choosing $n$ quarks is not $N^n$, but the much smaller  binomial coefficient
$\(\matrix{N\cr n}\)$.}. In order to match onto the effective nonrelativistic
operator \suivop\ we work to zeroth order in the baryon velocities; thus the
only source of angular momentum in the problem are the baryon spins.  The
spin-flavor dependence of a given diagram leading in $1/N$ is
 \eqn\interax{
N\,\bra{B_2}{\CO^{(n)}_{IJ}\over
N^n}\ket{B_1}\bra{B_4}{\bar\CO^{(n')}_{IJ}\over N^{n'}}
\ket{B_3}}
where the operator $\CO^{(n)}_{IJ}$ is an $n$-quark operator on the upper
baryon line with isospin and spin $(IJ)$, and $\bar\CO^{(n')}$ is an $n'$-quark
operator on the lower baryon line with the same $(IJ)$ and conjugate $I_3$,
$J_3$; the two operators are contracted so that the amplitude is a spin-isospin
singlet. Note that the spin of a quark is defined in the collision
center-of-momentum frame.

The first step in proving that the amplitudes in \interax\ imply $SU(4)$
symmetry is to prove that
\eqn\ijrule{\bra{B'}{\CO^{(n)}_{IJ}\over N^n}\ket{B} \ltap {1\over N^{|I-J|}}\
,}
for states $B,B'$ that have $I\sim J\sim\CO(1)$, and that operators that
saturate the bound \ijrule\ have a particular form.  Eq. \ijrule\  contains as
a specific case the $I_t=J_t$ rule discussed in the Skyrme model
\ref\itjt{M. P. Mattis, M. Mukherjee, \prl{61}{1988}{1344};
M. P. Mattis, E. Braaten, \physrev{D39}{1989}{2737};
N. Dorey, J. Hughes, M. P. Mattis, \prl{73}{1994}{1211};
 N. Dorey, M. P. Mattis, hep-ph/9412373
}.  The dominance of $I=J$ couplings is well known for the pion;  it also is
observed for the rho meson \ref\rhomeson{G.E. Brown, R. Machleidt,
\physrev{C50}{1994}{1731}}.   Eq. \ijrule\ was previously derived in ref.
\djmii.

Matrix elements such as in eq. \interax\ can be represented by strings of
one-quark operators acting on the baryon.  Each of these 1-quark operators can
have any of the 16 $SU(2)_I\times SU(2)_J$  quantum numbers appropriate for a
$q \bar q$ pair: $(0,0)$, $(1,0)$, $(0,1)$ or $(1,1)$, and can be represented
 as
\eqn\ibod{{\bf 1 } \ ,\quad {\bf I}_a \ ,\quad {\bf J}_i\ ,\quad {\bf G}_{ia}\
.}
The first three of these operators are simply the generators of quark number,
isospin, and spin respectively. Taken together, the
16 operators \ibod\   generate  $SU(4)\times U(1)$ symmetry, with commutators
of the generic form
\eqn\commu{  \[{\bf I, I}\] \sim {\bf I}\ ,\quad \[{\bf J,J}\]\sim{\bf J}\
,\quad \[{\bf I,G}\]\sim
\[{\bf J,G}\]\sim{\bf G}\ ,\quad \[{\bf G,G}\]\sim {\bf I+J}\ ,}
all other commutators vanishing.  Each of these operators when acting on a
baryon state $\ket{B}$ produces another baryon state $\ket{B'}$, since they do
not change quark number.  When $\ket {B}$  is a state with $I\sim J\sim
\CO(1)$, then the operators ${\bf  I }   $ and ${\bf  J }   $ produce
$\ket{B'}$ with
amplitude $\CO(1)$, while the operators ${\bf  1 }   $ and ${\bf  G }   $
produce $\ket{B'}$
with amplitude $\CO(N)$.   Thus the leading operators $\CO^{(n)}$ are of the
form
\eqn\leadop{\CO^{(n)}={\bf  G }^r{\bf 1}^{n-r}\ .}
Furthermore,    while  ${\bf  G}^2$ generically has a matrix element that is
$\CO(N^2)$, from the commutation relations \commu, the matrix element of
$\[{\bf  G}_{ia},{\bf  G}_{jb}\]$ has a matrix element $\CO(1)$.  Thus the
${\bf  G}$'s
in the leading operators \leadop\ are totally symmetrized.
Now consider how the indices of the ${\bf  G}$ operators might be contracted.
Using Fierz-type identities for Pauli matrices, one can show that \djmii
\eqn\fierzi{{\bf  G}_{ia} {\bf  G}_{ib} = \delta_{ab}\({\bf  1}{\bf  1} -{\bf
I}_c{\bf  I}_c\) +
{\bf  I}_a{\bf  I}_b + {\rm (1-body\ operators)}}
\eqn\fierzii{{\bf  G}_{ia} {\bf  G}_{jb}\epsilon_{abc}\epsilon_{ijk} = -2\({\bf
 G}_{kc}{\bf  1} -
{\bf  J}_k{\bf  I}_c\)  + {\rm (1-body\ operators)}}
and so the leading operators  can always be written in the form \leadop\ with
the $r$  ${\bf  G}$ operators totally symmetrized, and with none of the ${\bf
G}$
indices contracted.  Such an operator has $I=J=r$.   Similar arguments show
that the largest $n$-quark operator with $I=J+t$ has the form
${\bf  G}^J{\bf  I}^t{\bf  1}^{(n-J-t)}$ and has a matrix element of order
$N^{(n-t)}$.
This proves the assertion \ijrule.

The same identity \fierzii\ which allowed us to ignore as subleading any
antisymmetrized pair of spin or isospin indices allows us to independently
rearrange those indices on either of the baryon lines.  Thus we can write the
leading large-$N$  contributions to the baryon-baryon
interaction \interax\ in the form
\eqn\xform{N   \({1\over N^2}\sum_{i,a=1}^3 \{{\bf  G}_{ia}\}_1 \{{\bf
G}_{ia}\}_2+
\CO(1/N^2)\)^I}
times any number of identity operators, where $I$ equals the $t$-channel spin
and isospin;   the parentheses $\{\ \}_1$ and $\{\ \}_2$ refer to which of the
two baryon lines the 1-quark operators act upon.   (We omit the unit operators
which do not
modify the $SU(2)_J\otimes SU(2)_I$ structure).
The proof of $SU(4)$ invariance now follows trivially.  If we denote the
fifteen $SU(4)$ generators by $T_\mu = \{{\bf I}_a, {\bf J}_i,{\bf G}_{ia}\}$
and use the fact
that the matrix elements of $I_a$ and $J_i$ are $\CO(1)$, it follows that the
leading interaction \xform\ may be rewritten as
\eqn\suivinv{ N\( {1\over N^2} \sum_{\mu=1}^{15} \{{\bf  T_\mu}\}_1 \{{\bf
T_\mu}\}_2 +
\CO(1/N^2)\)^I}
which is manifestly $SU(4)$ invariant, up to corrections suppressed by $1/N^2$.
 This
shows that each individual connected graph contributing in Fig. 2,  summed over
colors,  is $SU(4)$ symmetric.

Similar arguments go through for three flavors, and one finds that in the
large-$N$ limit of QCD with $SU(3)$ flavor symmetry,  the low energy effective
theory will have $SU(6)$ invariant baryon interactions (for low $I,J,S$
states).   For equal mass quarks, corrections to $SU(6)$ are of order $1/N$,
however, rather than $1/N^2$ as in the two flavor case.  One source  of the
$1/N$ correction is that diagrams involving transfer of a strange quark between
baryons are subleading by $1/N$; another source is that the hypercharge matrix
$T_8$ is proportional to the unit operator plus $1/N$ corrections.  In addition
to the $1/N$ corrections, there will be $SU(6)$ violation proportional to
powers of $m_s$, the strange quark mass.  These corrections are potentially
quite important for phenomenology.

So far we have not discussed the space dependence of the contributions from
Fig. 2.
We assume that the only long-range contribution comes from pion exchange, and
that the remaining contributions can be Taylor expanded in terms of
$q^2/\Lambda^2$, ($\Lambda$ given roughly by the vector meson masses)
in the spirit of chiral perturbation theory.
We stress that the $SU(2f)$ invariance we found in the large-$N$ limit holds
separately for each connected contribution \interax.  This may seem
counter-intuitive:  $SU(4)$ symmetry in the quark model unites the $\pi$ and
the $\rho$ mesons into a single multiplet.  Since the $\rho$ and $\pi$ have
such different masses, $SU(4)$ is badly broken in the meson sector.  One might
conclude that since meson exchange plays a big role in nuclear interactions,
those interactions would have large $SU(4)$ violation as well. In fact, as far
as the $SU(4)$ symmetry in
nuclear forces goes, the $\pi$ and  $\rho$ are not united into the same
multiplet.  Rather each couples to the baryons with strength $N$ and coupling
$G_{ia}$  \itjt.  From the argument relating eqs. \xform\ and \suivinv, each
$\pi$ exchange and each $\rho$ exchange is independently $SU(4)$ invariant,
with corrections of $\CO(1/N^2)$.  One can think of the $\pi$ and the $\rho$ as
being in different, incomplete $SU(4)$ multiplets, where the missing members of
the multiplets would make negligible contributions to the nuclear force.
(This argument  only holds for $S$-wave interactions, for which $  \{q_i{\bf
G}_{ia}\}_1   \{q_j{\bf G}_{ja}\}_2\sim q^2 \{{\bf  G}_{ka}\}_1 \{{\bf
G}_{ka}\}_2$.)

\newsec{Conclusions}

We conclude with a  list of future directions for this line of
investigation.  One is to further the work of \refs{\weinberg-\sav}\  relating
the leading baryon interaction coefficients to the effective masses and
properties of baryons in matter.  In particular, one might analyze hypernuclei
and $\Sigma^-$ atoms in light of the predictions \resiii. One might
also apply the predictions
\resiii\ to the controversy of how strangeness first appears in dense matter --
in the form of kaon condensation, or in the form of hyperons \ref\nstar{See,
for example,
 R. Knorren, M. Prakash, P.J. Ellis,  nucl-th/9506016;
J. Schaffner, I. N. Mishustin, nucl-th/9506011 }.
However, these analyses require care: factors of $1/9$ multiplying $b$ in eq.
\resii\ are formally of order $1/N^2$.
Thus to leading order in $N$, $c_5=2a$ and the other $c_i$'s vanish giving rise
to an accidental $SU(16)$ symmetry in the low energy self-interactions of the
sixteen $J=1/2$ baryon octet states.  This $SU(16)$ symmetry is broken by
subleading $SU(6)$ violating operators suppressed by $1/N$ or $m_s$, which are
expected to be important phenomenologically  \ref\kapman{D.B. Kaplan, A.V.
Manohar, in preparation.}.   Among  the four nucleon isospin and spin states
there is an accidental $SU(4)_{sm}$ symmetry which is much more robust,
broken only at order $1/N^2$, and by isospin breaking.

Another suggestive line of inquiry is to extend the results of this Letter
beyond the dimension six operators \suivop.  In particular, one should
understand how $SU(2f)$ breaking comes about in the $L\cdot S$ interactions,
including analysis  of the cases where $SU(4)_{sm}$ seems to work better than
it should, in large nuclei where spin-orbit interactions are large --- most
notably the Franzini-Radicatti mass formula \ref\franrad{P. Franzini, L.A.
Radicatti, Phys. Lett. 13 (1963) 322}, \parikh.  Finally, it may prove fruitful
to apply $SU(4)$ symmetry to the problem of the quenching of Gamow-Teller
strengths in the shell \hbox{model \shellgt.}

\vfill\eject
\centerline{Acknowledgements}

We would like to thank G. Bertsch, H. Georgi, F. Gilman, W. Haxton,  R.
Kavanagh, L. Kisslinger, A. Manohar, R. Sorensen, U. van Kolck, M. Wise and L.
Wolfenstein for useful conversations and correspondence.
MJS would like to thank the INT for hospitality during a recent visit, where
this work was initiated.  MJS was supported in part by the U.S. Dept.
of Energy under Grant No.
DE-FG02-91-ER40682.
DK was supported in part by DOE grant DOE-ER-40561, NSF
Presidential Young Investigator award \pyidk, and by a grant from the
Sloan Foundation.

\listrefs

\bye